\documentclass[aps,prd,floatfix,nofootinbib,superscriptaddress,twocolumn]{revtex4-2}

\usepackage[T1]{fontenc}
\usepackage[utf8]{inputenc}
\usepackage{amsmath,amsfonts,amssymb}
\usepackage{graphicx}
\usepackage{bm}
\usepackage[normalem]{ulem}
\usepackage{xcolor}

\usepackage{array}
\usepackage[colorlinks=true,linkcolor=blue,citecolor=blue,urlcolor=blue]{hyperref}

\graphicspath{{figures/tikz/}}

\newcommand{\tr}{{\rm tr}}
\newcommand{\GeV}{{\rm GeV}}
\newcommand{\id}{\mathbf{1}}
\newcommand{\gQ}{\gamma^{Q}_{\mu\nu}}   
\newcommand{\gG}{\gamma^{G}_{\mu\nu}}   
\newcommand{\Xmu}{\mathcal{X}_\mu}

\begin{document}

\title{Pion gravitational form factors in an algebraic rainbow-ladder model}

\author{Jiayin Kang}
\affiliation{School of Physics, Nankai University, Tianjin 300071, China}

\author{Lei Chang}
\email[Corresponding author: ]{leichang@nankai.edu.cn}
\affiliation{School of Physics, Nankai University, Tianjin 300071, China}

\date{\today}

\begin{abstract}
Within rainbow-ladder truncation, we identify three contributions to the pion's gravitational form factors---graviton coupling to the quark, to the gluon dressing the quark, and to the gluon binding the pion---and separate them into quark and gluon parts.
Using an algebraic quark model with a dynamical basis, we obtain \(A(0)=1\), \(D(0)=-1\) in the chiral limit, and momentum fractions \(\langle x\rangle_q=0.896\), \(\langle x\rangle_g=0.104\).
The \(D\)-term obeys the same partition at the model scale.
\end{abstract}

\maketitle

\section{Introduction}\label{sec:intro}
Gravitational form factors provide access to the mass distribution and internal force densities inside hadrons, complementing the charge and current distributions encoded in electromagnetic form factors \cite{Kobzarev:1962wt,Pagels:1966zza,Polyakov:2018zvc,Burkert:2023wzr,Ji:2026lyj}.
As the lightest hadron, 
the pion occupies a privileged place in this programme \cite{Roberts:2021nhw}.
It is simultaneously a quark+antiquark bound-state and, in the chiral 
limit, a Nambu--Goldstone boson of QCD.
Its internal structure therefore carries the imprint of both confinement 
and dynamical chiral symmetry breaking \cite{Nambu:1961tp}, making the 
pion a minimal laboratory in which the two most distinctive features of 
QCD act together.
Moreover, in the chiral limit, symmetry imposes model-independent 
constraints on pion observables, which serve as a necessary benchmark: 
any calculation that fails to reproduce them is devoid of meaning.

Direct graviton scattering cannot be used to measure gravitational form factors. They can only be extracted indirectly \cite{Kumano:2017lhr}.
On the theoretical side, lattice-regularised QCD has provided results for the pion's gluon gravitational form factors \cite{Shanahan:2018pib} and for its gravitational form factors more generally \cite{Hackett:2023nkr,Bollweg:2026nnr}. Phenomenological models have 
also been used to study hadron gravitational form factors \cite{Broniowski:2008hx,Cao:2023ohj,Broniowski:2024oyk,Choi:2025rto,Deng:2026ydm,Liu:2026isp}.

Within continuum Schwinger function methods (CSMs), a dressed \(\pi\)--\(\pi\) amplitude has been proposed to calculate pion gravitational form factors with a contact interaction \cite{Xing:2022mvk}. This framework was subsequently extended to other pseudoscalar mesons \cite{Sultan:2024hep,Xing:2025uwn}.
CSMs calculations based on dressed quark--graviton vertices have delivered pion and kaon gravitational form factors \cite{Xu:2023izo}, with subsequent work also considering the nucleon \cite{Yao:2025fnb}.

The framework employed widely is the rainbow-ladder(RL) truncation, which is the leading term in a symmetry-preserving CSMs scheme \cite{Munczek:1994zz,Bender:1996bb}. In the chiral limit, it realizes the pion as a Nambu--Goldstone boson \cite{Maris:1997tm}.
Previous CSMs studies of the pion's gravitational form factors have focused on the graviton--quark coupling \cite{Xing:2022mvk,Xu:2023izo}, where the graviton is attached to the dressed-quark line, and the resulting contribution is closed by a dressed \(\pi\)--\(\pi\) amplitude, as in Fig.\,\ref{fig:diagrams}(a).
A graviton couples to all energy and momentum, including that carried by the interaction which dresses the quark and binds the pion. Since this interaction generates the dressed-quark mass function, coupling the graviton solely to the quark leaves the bound-state momentum incomplete. Recovering both the momentum sum rule and the chiral-limit soft-pion relation therefore requires that the graviton also be attached to the interaction.
The resulting contributions are depicted in Fig.~\ref{fig:diagrams}, where the graviton couples to the quark, to the gluon dressing the quark, and to the gluon binding the pion.
Given the distinct physical origins of these terms, we separate them into quark and gluon contributions: diagram (a) is the quark contribution, and diagrams (b) and (c) together form the gluon contribution.
This separation is defined at the level of the matrix element and can therefore be tested directly against the momentum sum rule and the soft-pion relation.

Evaluating these contributions requires a self-consistent numerical solution of the quark propagator, the Bethe--Salpeter amplitude, and the \(\pi\)--\(\pi\) amplitude, etc., within a given model interaction, \(\mathcal{D}\).
This entails substantial numerical challenges and error control.
The primary aim of this work is instead to provide a general computational framework and to demonstrate that momentum conservation and the soft-pion theorem hold exactly.
We therefore avoid the numerical complications at this stage and employ an algebraic model.
Such models express the dressed-quark propagator and pion Bethe--Salpeter amplitude in closed form while preserving the link between them, and have clarified many results (see, e.g., Refs.~\cite{Chang:2013pq,Chang:2016vkv,Sultan:2024mva}).
However, they specify no interaction, leaving the graviton couplings in Fig.~\ref{fig:diagrams}(b), (c) undetermined.
Here we construct such an interaction.
Following Ref.~\cite{Naito:1998sq}, we retain only the scalar and pseudoscalar interaction channels and build a separable interaction to match the quark mass function.
The algebraic propagator and pion amplitude then become exact solutions of the gap and Bethe--Salpeter equations, so that all contributions in Fig.~\ref{fig:diagrams} reduce to explicit integrals amenable to direct tests of the momentum sum rule and soft-pion relation.


This paper is organised as follows.
Section\,\ref{sec:rl} describes the RL framework and the three graviton contributions to the pion's form factors.
Section\,\ref{sec:model} develops the algebraic model with a dynamical basis, and Sec.\,\ref{sec:features} presents the graviton couplings it induces.
Results are reported in Sec.\,\ref{sec:results}, where we discuss the momentum sum rule, the soft-pion relation, and the channel-resolved partition of \(A\) and \(D\).
Finally, Sec.\,\ref{sec:summary} offers a summary and perspective.

\section{Gravitational form factors in rainbow-ladder truncation}\label{sec:rl}
We adopt the quark gravitational form factor as defined in Ref.~\cite{Xing:2022mvk}, which can be written as\footnote{We work in Euclidean space, where $\{\gamma_\mu,\gamma_\nu\}=2\delta_{\mu\nu}$ and $\int_q\equiv\int d^4q/(2\pi)^4$.
All external momenta are incoming: the pion momenta are $P_1$ and $P_2$, while the graviton momentum is $Q=-(P_1+P_2)$.
With $P=\tfrac12(P_1-P_2)$, one has $P\cdot Q=0$ and $P_1^2=P_2^2=-m_\pi^2$.},
\begin{align}
\mathcal M^{(a)}_{\mu\nu}={}&2\times2N_c\,\tr\!\int_q\gQ(q-P_2,q+P_1)\,S(q+P_1)\nonumber\\
&\times i^2F(q,P_1,P_2)\,S(q-P_2)\,,\label{eq:Maint}
\end{align}

\begin{figure*}[t]
\centering\includegraphics{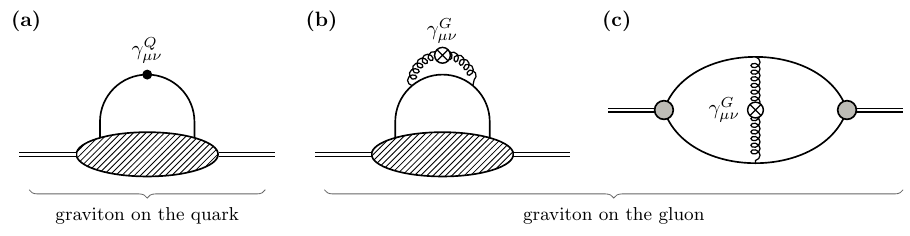}
\caption{\label{fig:diagrams}
Contributions to the pion's $A$ and $D$ form factors in RL truncation, Eq.\,\eqref{eq:three}.
Graviton lines are suppressed; a black dot marks the quark--graviton vertex, $\gQ$, and a crossed circle marks the gluon--graviton vertex, $\gG$.
(a) The graviton couples to the quark; the diagram is closed with the $\pi$--$\pi$ scattering amplitude, $F$ (hatched ellipse).
(b) The graviton couples to the gluon that dresses the quark; the diagram is closed with $F$.
(c) The graviton couples to the gluon that binds the pion.
Double straight lines: pions; single lines: dressed quarks; curly lines: the RL interaction $\mathcal D_{\alpha\beta}$; grey discs: pion Bethe--Salpeter amplitudes.}
\end{figure*}

In the calculation, we introduced the bare quark-graviton interaction vertex,
\begin{equation}
\gQ(k_+,k_-)=i\gamma_\mu k_\nu-\tfrac12\delta_{\mu\nu}\big[S_0^{-1}(k_+)+S_0^{-1}(k_-)\big]\,,\label{eq:gQ}
\end{equation}
where $k_+=q-P_2$, $k_-=q+P_1$, and the bare quark propagator $S_{0}^{-1}(k)=i\gamma\cdot k+m$; the leading factor of two includes the reflected contribution, in which the graviton couples to the antiquark.

In addition, the energy-momentum tensor involves the dressed-quark propagator \(S\), the bound-state Bethe--Salpeter amplitude \(\Gamma_{\mathcal{M}}\), and the quark-quark correlation function \(F\).
Within the RL truncation, these ingredients are determined by a common kernel in a symmetry-preserving manner \cite{Maris:1997tm,Xu:2023izo}:
\begin{equation}
\mathcal K^{rs}_{tu}(k,q)=-\frac43\,\mathcal D_{\alpha\beta}(k-q)\,[\gamma_\alpha]_{ts}\,[\gamma_\beta]_{ru}\,,
\label{eq:kernel}
\end{equation}
where \(r\), \(s\), \(t\), and \(u\) label Dirac indices, repeated indices are summed, and the color factor is included in the prefactor.
The RL interaction, \(\mathcal D_{\alpha\beta}(\ell)\), combines the dressed-gluon propagator and quark--gluon-vertex dressing; see, e.g., Refs.~\cite{Maris:1999nt,Qin:2011dd}.

First, the quark propagator is determined by the rainbow gap equation:
\begin{equation}
[S^{-1}(k)]_{tu}=[S_0^{-1}(k)]_{tu}-\int_q[S(q)]_{sr}\,\mathcal K^{rs}_{tu}(k,q)\,.
\label{eq:gap}
\end{equation}
Here we omit the renormalization-related factors, which can be readily incorporated within the RL truncation.

Second, the matched ladder Bethe--Salpeter equation fixes the meson amplitude:
\begin{equation}
[\Gamma_\mathcal{M}(k;P)]_{tu}=\int_q\big[\chi_{\mathcal{M}}(q;P)\big]_{sr}\mathcal K^{rs}_{tu}(k,q)\,,
\label{eq:bse}
\end{equation}
where \(\chi_{\mathcal{M}}(q;P)=S(q+\tfrac P2)\,\Gamma_\mathcal{M}(q;P)\,S(q-\tfrac P2)\).

Finally, the general quark-quark correlation function appearing in Fig.~\ref{fig:diagrams}(a) obeys the ladder equation
\begin{equation}
[F(k,P_1,P_2)]_{tu}=[F_0]_{tu}+\int_q\big[\mathcal{F}(q,P_1,P_2)\big]_{sr}\mathcal K^{rs}_{tu}(k,q)\,,
\label{eq:F}
\end{equation}
with
\begin{equation}
F_0=\Gamma_\mathcal{M}(k+\tfrac{P_1}{2};P_1)S(k)\Gamma_\mathcal{M}(k-\tfrac{P_2}{2};P_2)
\label{eq:F0}
\end{equation}
and
\begin{equation}
\mathcal{F}(q,P_1,P_2)=S(q+P_1)\,F(q,P_1,P_2)\,S(q-P_2)\,.
\label{eq:F01}
\end{equation}

The Born term, \(F_0\), connects the two meson Bethe--Salpeter vertices through the quark line \(S(k)\), each evaluated at the relative momentum of the corresponding quark pair.
The second term on the right-hand side of Eq.~\eqref{eq:F} resums ladder exchanges between the quark legs \(S(q+P_1)\) and \(S(q-P_2)\).
Hence diagram (a) involves the full \(F\), not merely its Born term \(F_0\).
This dressed \(\pi\)--\(\pi\) rescattering mechanism was employed with a contact interaction in Ref.~\cite{Xing:2022mvk} and extended to other pseudoscalar mesons in Ref.~\cite{Sultan:2024hep}.

The quark in Fig.~\ref{fig:diagrams}(a) is dressed by the interaction in Eq.~\eqref{eq:gap}, which also binds the quarks through Eq.~\eqref{eq:bse} to form the bound state.
Since the EMT includes the energy and momentum carried by this interaction, the graviton probes both the interaction and the quark.
Consequently, in the model below, coupling the graviton only to the quark does not exhaust the bound-state momentum.
We therefore introduce two additional contributions.
Diagram (b) depicts the graviton coupled to the gluon that dresses the quark, while diagram (c) shows its coupling to the gluon that binds the bound state.
Given the distinct origins of these terms, we define the quark and gluon contributions separately \footnote{Ref.\,\cite{Xu:2023izo} has already pointed out the necessity of this contribution, but has not explicitly provided a specific form},
\begin{equation}
\begin{aligned}
\mathcal M^{q}_{\mu\nu}&=\mathcal M^{(a)}_{\mu\nu}\,,\quad
\mathcal M^{g}_{\mu\nu}=\mathcal M^{(b)}_{\mu\nu}+\mathcal M^{(c)}_{\mu\nu}\,.
\end{aligned}
\label{eq:three}
\end{equation}


Attaching the graviton to the gluon in the gap kernel amounts to the replacement $\mathcal D_{\alpha\beta}\to-[\mathcal D\,\gG\,\mathcal D]_{\alpha\beta}$; this gives the insertion in diagram (b), where $\ell_\pm=k_\pm-p$ are the gluon momenta:
\begin{subequations}\label{eq:Mb}
\begin{equation}
\gQ(k_+,k_-)\to-\int_p[S(p)]_{sr}\,\mathcal K^{rs}_{\mu\nu;tu}(k_+,k_-;p)\,,\label{eq:Mbrep}
\end{equation}
with
\begin{equation}
\begin{split}
\mathcal K^{rs}_{\mu\nu;tu}(k_+,k_-;p)={}&\big[\mathcal D(\ell_+)\,\gG(\ell_+,\ell_-)\,\mathcal D(\ell_-)\big]_{\alpha\beta}\\
&\times\tfrac43\,[\gamma_\alpha]_{ts}\,[\gamma_\beta]_{ru}\,.
\end{split}\label{eq:Kmunu}
\end{equation}
\end{subequations}
For diagram (c), the same graviton attachment is made to the kernel in Eq.\,\eqref{eq:bse}, whose explicit expression in the model is given in Eq.\,\eqref{eq:Mc}.

For the effective gluon--graviton vertex common to diagrams (b) and (c), we impose Ward--Green--Takahashi identity (WGTI) \cite{Brout:1966oea},
\begin{equation}
Q_\mu\gamma^G_{\mu\nu;\alpha\beta}(\ell_+,\ell_-)=\mathcal D^{-1}_{\alpha\beta}(\ell_+)\ell_{-\nu}-\mathcal D^{-1}_{\alpha\beta}(\ell_-)\ell_{+\nu}\,,\label{eq:wgtiG}
\end{equation}
where $\ell_\pm$ are the gluon momenta before and after graviton absorption; $\alpha$$\beta$ label gluon Lorentz indices, and $\mathcal D^{-1}_{\alpha\beta}$ denotes the inverse of $\mathcal D_{\alpha\beta}$ in these indices.
%
%
The right-hand side reduces to $[\mathcal D^{-1}_{\alpha\beta}(\ell_+)-\mathcal D^{-1}_{\alpha\beta}(\ell_-)]\ell_\nu-Q_\nu\frac{\mathcal D^{-1}_{\alpha\beta}(\ell_+)+\mathcal D^{-1}_{\alpha\beta}(\ell_-)}{2}$, with $\ell=\tfrac12(\ell_++\ell_-)$, and the effective gluon-graviton vertex can be solved properly.

In summary, under the RL truncation and with a model-dependent interaction input, we have constructed a procedure for calculating the gravitational form factor, which can be applied to the calculation of two-body bound-states with arbitrary quantum numbers. In the following, we will take the pion as an example to discuss whether this framework can reproduce momentum conservation and the soft-pion limit in the chiral limit.

For each channel, $j=q,g$, the pion-state matrix element takes the form
\begin{align}
\mathcal M^j_{\mu\nu}={}&2P_\mu P_\nu A_j(Q^2)+\tfrac12\big(Q_\mu Q_\nu-Q^2\delta_{\mu\nu}\big)D_j(Q^2)\nonumber\\
&-2m_\pi^2\delta_{\mu\nu}\,\bar c_j(Q^2)\,,
\label{eq:decomp}
\end{align}
where $A=A_q+A_g$, $D=D_q+D_g$, and $\bar c_q+\bar c_g=0$.
On shell, $A_j$ and $D_j$ are isolated using the projectors
\begin{subequations}\label{eq:proj}
\begin{align}
\mathcal{P}^{A}_{\mu\nu}&=\frac{1}{4P^2}\Big(\frac{3P_\mu P_\nu}{P^2}+\frac{Q_\mu Q_\nu}{Q^2}-\delta_{\mu\nu}\Big)\,,\\
\mathcal{P}^{D}_{\mu\nu}&=-\frac{1}{Q^2}\Big(-\frac{P_\mu P_\nu}{P^2}-\frac{3Q_\mu Q_\nu}{Q^2}+\delta_{\mu\nu}\Big)\,.
\end{align}
\end{subequations}
These two projectors are orthogonal to \(\delta_{\mu\nu}\), extracting \(A_j\) and \(D_j\), respectively; terms proportional to \(\delta_{\mu\nu}\) in matrix elements or vertices are therefore omitted.

It is important to note that energy-momentum conservation and the soft-pion theorem impose the following two constraints:
\begin{subequations}\label{eq:constraints}
\begin{align}
\sum_{j}A_{j}(0)&=1\,,\label{eq:A1}\\
\sum_{j}A_j(0)+D_j(0)&\overset{m_\pi=0}{=}0\,.\label{eq:softpion}
\end{align}
\end{subequations}
Any calculation must satisfy these constraints from the outset.

\section{Algebraic model with a dynamical basis}\label{sec:model}

For a specified interaction, evaluating these contributions requires solving the gap equation at complex momenta, the Bethe--Salpeter equation at both on-shell pion momenta, and Eq.~\eqref{eq:F} as a four-variable integral equation; diagram (c) also requires a two-loop integral and the construction of \(\gG\).
Since numerical errors accumulate at each stage, it is preferable to first employ a model in which every stage can be completed with negligible numerical error, thereby providing a benchmark for the general calculation.

Consider the quark propagator and pion amplitude \cite{Sultan:2024mva}:
\begin{equation}
\begin{aligned}
S^{-1}(k)&=i\gamma\cdot k +B(k^2)\,,\\
B(k^2)&=\frac{M^3}{k^2+M^2}\,,\\
\Gamma_\pi(k;P)&=i\gamma_5\,\frac{B(k^2)}{f_\pi}\,,
\end{aligned}
\label{eq:alg}
\end{equation}
Pion amplitude $\Gamma_\pi$ is constructed from the Goldberger--Treiman relation\cite{Maris:1997tm} and $k$ is the quark+antiquark relative momentum; the parameter $M=0.50551\,\GeV$ has been fitted by reproducing the chiral-limit decay constant $f_\pi=0.0924\,\GeV$ to numerical accuracy.
The mass function approaches $B(0)=M$ in the infrared and falls as $1/k^2$ in the ultraviolet, as expected for the dressed-quark mass function.
Related algebraic models have been used to study the pion light-front wave function and distribution amplitude \cite{Chang:2013pq}; the pion distribution function and its elastic and transition form factors \cite{Chang:2016vkv}; and the $\gamma^\ast\pi^0\gamma$ transition form factor \cite{Sultan:2024mva}.


Unlike the calculation of electromagnetic form factors or distribution amplitudes, the gravitational form factors require more than a propagator and a pion amplitude: diagrams (b) and (c) also involve the interaction \(\mathcal D\), through the graviton-attached kernel in Eq.~\eqref{eq:Kmunu} and the constraint in Eq.~\eqref{eq:wgtiG}.
Since \(\mathcal D\) must be compatible with the propagator and pion amplitude, we seek an interaction and truncation in which the algebraic forms in Eq.~\eqref{eq:alg} solve both the gap and Bethe--Salpeter equations exactly.
We proceed in two stages: first the truncation and separable interaction, then the resulting Bethe--Salpeter and scattering amplitudes.

First, we need a truncation scheme in which only the scalar part of the inverse quark propagator is dressed, i.e.\ \(A(k^2)\equiv 1\).
Following Ref.~\cite{Naito:1998sq}, a self-consistent scheme exists, in which the interaction kernel takes the form
\begin{equation}
    \mathcal{K}^{rs}_{tu}(k,q)=-\tfrac43\,\mathcal D(k-q)\frac{1}{4}\left\{\big[\id\big]_{tu}\big[\id\big]_{rs}+\big[i\gamma_{5}\big]_{tu}\big[i\gamma_{5}\big]_{rs}\right\}\,. \label{eq:fierz}
\end{equation}
Accordingly, the gap equation, Eq.~\eqref{eq:gap}, becomes
\begin{equation}
A(k^2)\equiv1\,,\quad
B(k^2)=\frac43\int_q\mathcal D(k-q)\,\sigma_S(q^2)\,,
\label{eq:gapfierz}
\end{equation}
with \(\sigma_S(q^2)=B(q^2)/[q^2+B^2(q^2)]\) in the chiral limit.

To obtain a tractable kernel, we first expand it in \(O(4)\) hyperspherical harmonics and then adopt a separable expansion for the leading term \cite{Cahill:1995ka}:
\begin{equation}
\mathcal D(k-q)\to G(k^2)\,G(q^2)\,.
\label{eq:sep}
\end{equation}
Equation~\eqref{eq:gapfierz} then becomes \(B(k^2)=\tfrac43G(k^2)\int_qG(q^2)\sigma_S(q^2)\); hence \(B\propto G\), and, up to an overall sign,
\begin{equation}
G(k^2)=\frac{B(k^2)}{b_1}\,,\qquad b_1^2=\frac43\int_q B(q^2)\,\sigma_S(q^2)\,,
\label{eq:Gdef}
\end{equation}
where \(b_1\) is a constant.
Provided the integral in Eq.~\eqref{eq:Gdef} is finite, the separable interaction is fixed by this construction, except for an overall sign.
Notably, the propagator in Eq.~\eqref{eq:alg} solves Eqs.~\eqref{eq:gapfierz}--\eqref{eq:Gdef} exactly, which ensures that the model is dynamically motivated rather than purely algebraic.
The only parameter involved is \(M\).
Separable kernels have appeared previously \cite{Burden:1996nh,Blaschke:2000gd}; as in the present work, Ref.~\cite{Burden:1996nh} constructed its kernel from the quark propagator.

Meanwhile, the Bethe--Salpeter equation, Eq.\,\eqref{eq:bse}, becomes
\begin{align}
\Gamma_\pi(k;P)={}&-\frac43\,i\gamma_5\,G(k^2)\int_qG(q^2)\nonumber\\
&\times\frac14\tr\big[i\gamma_5\,S(q+\tfrac P2)\,\Gamma_\pi(q;P)\,S(q-\tfrac P2)\big]\,.
\label{eq:bsesep}
\end{align}
Only the $i\gamma_5$ amplitude remains; hence $\Gamma_\pi\propto i\gamma_5G(k^2)\propto i\gamma_5B(k^2)$ for any $P$, whilst $P^2$ affects only the normalisation.
At \(P^2=0\), the BSE reduces to the eigenvalue condition
\begin{equation}
\lambda=\frac{4}{3b_1^2}\int_q\frac{B^2(q^2)}{q^2+B^2(q^2)}=1\,,
\label{eq:lambda}
\end{equation}
which fixes \(b_1\).
The gap equation and the BSE yield the same condition, ensuring self-consistency and a massless pion in the chiral limit. The solution is $\Gamma_\pi(k;P)=i\gamma_5B(k^2)/f_\pi$, fixed by the canonical normali-sation condition~\cite{Maris:1997tm}.
Consequently, the amplitude in Eq.\,\eqref{eq:alg} is a solution of Eq.\,\eqref{eq:bse}, rather than an additional assumption.

Writing the amplitude as a Born term plus a self-energy, Eq.\,\eqref{eq:F} becomes
\begin{subequations}\label{eq:Fsigma}
\begin{align}
&F(k,P_1,P_2)=F_0(k,P_1,P_2)+\Sigma(k,P_1,P_2)\,,\label{eq:FF0S}\\
&[\Sigma(k,P_1,P_2)]_{tu}=\int_q\big[S(q+P_1)\,\{F_0(q,P_1,P_2)\nonumber\\
&\quad+\Sigma(q,P_1,P_2)\}\,S(q-P_2)\big]_{sr}\,\mathcal K^{rs}_{tu}(k,q)\,.\label{eq:Sigma}
\end{align}
\end{subequations}
The model kernel connects the quark pairs $(k+P_1,k-P_2)$ and $(q+P_1,q-P_2)$, whose relative momenta are $k+P$ and $q+P$.
Using Eq.\,\eqref{eq:sep}, so that $\mathcal D\to G((k+P)^2)\,G((q+P)^2)$, together with Eq.\,\eqref{eq:fierz}, we obtain
\begin{align}
\Sigma(k,P_1,P_2)={}&-\frac43\,G((k+P)^2)\int_qG((q+P)^2)\nonumber\\
&\times\frac14\Big\{\tr X(q)+i\gamma_5\,\tr\big[i\gamma_5X(q)\big]\Big\}\,,
\label{eq:ampfierz}
\end{align}
where
\begin{equation*}
X(q)=S(q+P_1)\big[F_0(q,P_1,P_2)+\Sigma(q,P_1,P_2)\big]S(q-P_2)\,.
\end{equation*}
This self-energy has the following properties.

(\emph{i}) The matrix \(X\) depends only on the three independent momenta \(q\), \(P_1\), and \(P_2\); hence its pseudoscalar trace vanishes identically.
Consequently, \(\Sigma\propto\id\), so the ladder acts solely in the scalar channel.

(\emph{ii}) Eq.\,\eqref{eq:ampfierz} implies $\Sigma(k,P_1,P_2)=G((k+P)^2)\,s$, where $s$ obeys $s=s_0+\lambda_\sigma s$; hence
\begin{subequations}\label{eq:s}
\begin{equation}
s=\frac{s_0}{1-\lambda_\sigma(Q^2)}\,,\label{eq:sclosed}
\end{equation}
with
\begin{equation}
\begin{split}
s_0={}&-\frac43\int_qG((q+P)^2)\\
&\times\frac14\tr\big[S(q+P_1)F_0(q,P_1,P_2)S(q-P_2)\big]\,,
\end{split}\label{eq:s0}
\end{equation}
and
\begin{equation}
\begin{split}
\lambda_\sigma(Q^2)={}&-\frac43\int_qG^2((q+P)^2)\\
&\times\frac14\tr\big[S(q+P_1)S(q-P_2)\big]\,.
\end{split}\label{eq:lsig}
\end{equation}
\end{subequations}

(\emph{iii}) The scalar-channel Bethe--Salpeter equation has eigenvalue $\lambda_\sigma$; accordingly, $\Sigma$ represents scalar exchange in the $t$-channel and develops a pole at $\lambda_\sigma=1$.
At $Q^2=0$ in the chiral limit, Eqs.\,\eqref{eq:lambda} and \eqref{eq:s} give
\begin{equation}
1-\lambda_\sigma(0)=\frac{8}{3b_1^2}\int_qB^2(q^2)\,\sigma_S^2(q^2)=0.325\,,
\label{eq:sigmafactor}
\end{equation}
\emph{i.e}.\ resummation enhances the single-rung term $s_0$ by a factor of $3.07$. Consequently, we obtain
\begin{equation}
\Sigma(k,P_1,P_2)=\frac{1}{2{f_{\pi}}^2}B((k+P)^2)\,.
\end{equation}

With the model fully specified, all ingredients entering Fig.~\ref{fig:diagrams} are now known.
We turn next to the graviton couplings induced in this model.
\section{Gluon--graviton couplings}\label{sec:features}

To evaluate diagrams (b) and (c), we need the effective gluon--graviton vertex.
In this model the effective gluon interaction is \(G(k^{2})G(q^{2})\), as specified in Eq.~\eqref{eq:sep}, and the graviton couples to either factor \(G\).
We impose the same Ward--Green--Takahashi identity as in Eq.~\eqref{eq:wgtiG}.
After omitting the \(\delta_{\mu\nu}\) terms, as before, and using \(B=b_{1}G\) from Eq.~\eqref{eq:Gdef}, the analogue of Eq.~\eqref{eq:wgtiG} is
\begin{equation}
Q_\mu\Xmu(k;Q)=B(k_+^2)-B(k_-^2)\,.
\label{eq:wgtiX}
\end{equation}
We use the straight solution~\cite{Munczek:1994zz}:
\begin{equation}
\Xmu(k;Q)=\int_{-1}^{1}\!dz\,\Big(k+\frac z2 Q\Big)_{\!\mu}\,
B'\!\Big(\big(k+\tfrac z2Q\big)^2\Big)\,.
\label{eq:Xmu}
\end{equation}
Here \(B'(t)=dB/dt\).
At \(Q=0\), Eq.~\eqref{eq:Xmu} becomes \(2k_\mu B'(k^2)\), the scalar part of the Ward-identity vertex \(\partial_{k_\mu}S^{-1}(k)=i\gamma_\mu+2k_\mu B'(k^2)\) \cite{Ward:1950xp}.

\emph{Diagram (b)}.
Attaching the graviton to the gluon in \(B(k^2)=\tfrac43G(k^2)\int_pG(p^2)\sigma_S(p^2)\) gives Eq.~\eqref{eq:Maint}, apart from \(\delta_{\mu\nu}\) terms, with \(\gQ\) replaced by
\begin{subequations}\label{eq:XY}
\begin{align}
&\Xmu(k;Q)\,k_\nu+B(k^2)\,Y_{\mu\nu}(Q)\,,\label{eq:Lmodel}\\
&Y_{\mu\nu}(Q)=\frac{\int_p\Xmu(p;Q)\,p_\nu\,\sigma_S(p^2)}{\int_pB(p^2)\,\sigma_S(p^2)}\,,
\label{eq:Y}
\end{align}
\end{subequations}
where \(k=q+P\).

\emph{Diagram (c)}.
The same graviton attachment is made to the kernel in Eq.~\eqref{eq:bsesep}, which reduces the two-loop contribution to a single loop:
\begin{align}
\mathcal M^{(c)}_{\mu\nu}={}&2\times2N_c\,\tr\!\int_q q_\nu\,i\gamma_5\,\frac{\Xmu(q+\tfrac{P_1}{2};Q)}{f_\pi}\nonumber\\
&\times S(q+P_1)\,\Gamma_\pi(q+\tfrac{P_1}{2};P_1)\,S(q)\,.
\label{eq:Mc}
\end{align}
The factor of two accounts for the two equal attachments, one to each factor in \(G\,G\).
Both diagrams (b) and (c) therefore contain the same \(\Xmu\).

With \( \Xmu\) determined, diagrams (b) and (c) reduce to explicit integrals.
We now present the resulting form factors.
\section{Results and discussion}\label{sec:results}

In the model employed here, all integral components can be cast into a propagator-like form, making the Feynman parametrisation both natural and straightforward. The value at \(Q^{2}=0\) is therefore obtained without numerical difficulty.

The individual contributions of the diagrams in Fig.\,\ref{fig:diagrams} are
\begin{equation}
\begin{array}{lcccc}
\text{diagram} & \text{vertex} & \text{channel} & A(0) & D(0) \\[2pt]
\text{(a)} & \gamma^Q & q & 0.896 & -0.896 \\
\text{(b)} & \gamma^G & g & 0.063 & 0.005 \\
\text{(c)} & \gamma^G & g & 0.041 & -0.109
\end{array}
\label{eq:table}
\end{equation}
with (b)\,$+$\,(c) giving \(A_g(0)=0.104\), \(D_g(0)=-0.104\).
Their sum gives
\begin{equation}
A(0)=1\,,\qquad D(0)=-1\,,
\label{eq:sumrule}
\end{equation}
satisfying the momentum sum rule, Eq.\,\eqref{eq:A1}, and the soft-pion theorem, Eq.\,\eqref{eq:softpion}.

Diagram (a) satisfies \(A+D=0\) on its own.
Neither diagram (b) nor (c) does so separately: they contribute with opposite signs and cancel only in sum.
Thus the channel-resolved relation, Eq.\,\eqref{eq:softpion}, is realised in this model.

The two components of \(F\) play different roles in diagram (a): \(F_0\) supplies all of \(A^{(a)}\), whereas \(\Sigma\) supplies \(69\%\) of \(D^{(a)}\) and nothing to \(A^{(a)}\).
Without diagram (c), one would obtain  \(A(0) = 0.959\).
We also verified, to within \(10^{-5}\), that the Bethe--Salpeter eigenvalue in Eq.\,\eqref{eq:lambda} is unity and that \(A(0)\) agrees with the pion electromagnetic form factor calculated with the same \(F\).

The resulting channel partition is
\begin{equation}
\begin{array}{ll}
\langle x\rangle_q=A_q(0)=0.896\,, & D_q(0)=-0.896\,,\\[2pt]
\langle x\rangle_g=A_g(0)=0.104\,, & D_g(0)=-0.104\,.
\end{array}
\label{eq:chanres}
\end{equation}
Notably, \(D(0)=-A(0)\) holds for each parton type.

It is useful to compare these results with previous CSM studies.
Reference~\cite{Xing:2022mvk} employed a contact interaction and a dressed \(\pi\)--\(\pi\) amplitude, but retained only the graviton--quark coupling; the gluon contributions identified here were not included.
Reference~\cite{Xu:2023izo} emphasized the need for a graviton--gluon coupling but did not provide an explicit form for it.
The present model supplies that form and shows that the gluon contributions are essential: without diagram (c), one would obtain \(A(0)=0.959\), violating the momentum sum rule.
Thus the framework developed here extends earlier treatments by making the gluon contributions explicit and by demonstrating that they are required for consistency.




\section{Summary and Perspective}\label{sec:summary}
Working within RL truncation, we identified the three contributions to the pion's \(A\) and \(D\) form factors: the graviton couples to the quark, to the gluon dressing the quark, and to the gluon binding the pion [Fig.~\ref{fig:diagrams}].
For the effective gluon interaction, we imposed a gluon--graviton identity of the same form as that for the quark.
We then constructed an algebraic quark model with a dynamical basis.
Retaining the scalar and pseudoscalar channels, together with a separable interaction constructed from the quark mass function, made its propagator and pion amplitude exact solutions of the gap and Bethe--Salpeter equations.
All contributions were consequently reduced to explicit integrals.

The calculation gave \(A(0)=1\) and \(D(0)=-1\).
The graviton--quark contribution obeyed the soft-pion relation independently, while the two graviton--gluon contributions obeyed it only in sum.
Within the \(\pi\)--\(\pi\) amplitude, ladder resummation represented scalar exchange in the \(t\)-channel.
It contributed nothing to \(A\) and most of \(D\).

Two clarifications are in order.
First, the ``gluon'' here denotes the effective interaction in RL truncation rather than a gauge-invariant gluon field.
Second, the model scale is undetermined, so comparison with lattice results at \(\zeta=2\,\GeV\) is premature \cite{Shanahan:2018pib,Hackett:2023nkr}.

A distinctive feature of this model is that the gluon propagator, the quark mass function, the pion Bethe--Salpeter amplitude, the \(\pi\)--\(\pi\) scattering amplitude, and the graviton--gluon vertex are all described in terms of a single, common function.
In principle, one may vary its functional form---for instance, by introducing a momentum-dependent mass function into the gluon propagator---thereby opening up a different and potentially more intriguing direction for further exploration.
Beyond the \(Q^2=0\) values considered here, the same setup can also give \(A_j(Q^2)\) and \(D_j(Q^2)\), and hence channel-separated mass and mechanical radii.
Restoring the \(\delta_{\mu\nu}\) structures omitted in this paper would enable tests of \(\bar c_q+\bar c_g=0\).
Finally, the prescription in Sec.~\ref{sec:rl} carries over to a non-separable RL interaction.
This work is in progress.

\begin{acknowledgments}
The authors are grateful to Zanbin Xing for constructive discussions. Work supported by: National Natural Science Foundation of China (Grant No. 12135007).
\end{acknowledgments}

\bibliography{refs}
\bibliographystyle{apsrev4-2}

\end{document}